\documentclass[english,aps,manuscript]{revtex4}
\usepackage[T1]{fontenc}
\usepackage[latin1]{inputenc}

\makeatletter

\usepackage{graphicx}

\usepackage{babel}

\usepackage{babel}
\makeatother
\begin{document}

\title{Surface Wetting of Liquid Nanodroplets: Droplet Size Effects}

\author{David R. Heine, Gary S. Grest, Edmund B. Webb III}

\affiliation{Sandia National Laboratories, Albuquerque, New Mexico 87185}

\date{\today{}}

\begin{abstract}
The spreading of liquid nanodroplets of different initial radii $R_{0}$
is studied using molecular dynamics simulation. Results for two distinct
systems, Pb on Cu(111), which is non-wetting, and a coarse grained
polymer model, which wets the surface, are presented for Pb droplets
ranging in size from $\sim55\,000$ to $220\,000$ atoms and polymer
droplets ranging in size from $\sim200\,000$ to $780\,000$ monomers.
In both cases, a precursor foot precedes the spreading of the main
droplet. This precursor foot spreads as $r_{f}^{2}(t)=2D_{eff}t$
with an effective diffusion constant that exhibits a droplet size
dependence $D_{eff}\sim R_{0}^{1/2}$. The radius of the main droplet
$r_{b}(t)\sim R_{0}^{4/5}$ in agreement with kinetic models for the
cylindrical geometry studied. 
\end{abstract}
\maketitle
Numerous practical problems are determined by the wetting of surfaces
by liquids including adhesion, lubrication, coatings, plant protection,
and oil recovery. The spreading of a droplet on a surface proceeds
as the droplet balances the interfacial tensions between the solid,
liquid, and vapor phases. Completely wetting droplets are known to
form a monolayer film or precursor foot on the surface that spreads
ahead of the droplet. For {}``high-energy'' surfaces, spreading
includes the formation of molecule-sized terraces along with the precursor
foot \cite{HCL:PRL:89,HCL:PRL:90}.

Although the spreading behavior of droplets on surfaces has been extensively
studied, droplet size effects on the spreading dynamics are often
ignored. These size effects can have a large influence on the wetting
behavior, particularly when using nanoscale droplets such as in MEMS
and microfluidic devices. For these devices, droplet size can strongly
affect both the manufacturing speed, such as in microcontact printing,
and the device performance.

For a spreading liquid droplet, the energy dissipation mechanics have
been classified by de Gennes. The total dissipation in the spreading
droplet can be expressed as a sum of three distinct dissipation mechanisms,
$T\Sigma_{w}+T\Sigma_{f}+T\Sigma_{l}$ \cite{G:RMP:85}. In this equation,
$T\Sigma_{w}$ is the contribution due to viscous dissipation in the
bulk of the droplet, $T\Sigma_{f}$ is the contribution due to viscous
dissipation in the precursor foot, and $T\Sigma_{l}$ is the contribution
due to adsorption of liquid molecules to the surface at the contact
line. The adsorption mechanism is expected to dominate for low viscosity
systems at short times while the bulk viscous dissipation mechanism
takes over at later times \cite{BG:ACI:92}. The dissipation from
the foot is less understood, though in the simulations presented here
we will show that this dissipation does not play a role in the spreading
of the droplet.

The size dependence of the bulk droplet spreading rate is present
in models of droplet spreading. In the molecular kinetic theory of
liquids \cite{GLE:TRP:41,BH:JCI:69}, the energy dissipation occurs
at the contact line. In the linearized version of this model, the
contact radius $r_{b}(t)$ of the bulk region of the droplet scales
with the droplet volume according to $r_{b}(t)\sim R_{0}^{6/7}$ at
late times for spherical droplets \cite{RCO:Lan:99} and $r_{b}(t)\sim R_{0}^{4/5}$
for cylindrical droplets \cite{HGW:PRE:04}. The hydrodynamic model
\cite{C:JFM:86,SB:JAP:94} is based upon the solution of the equations
of motion and continuity for the droplet and assumes that energy dissipation
occurs via viscous dissipation in the bulk. For the hydrodynamic model,
the contact radius of the bulk region scales according to $r_{b}(t)\sim R_{0}^{9/10}$
at late times for spherical droplets \cite{RCO:Lan:99} and $r_{b}(t)\sim R_{0}^{6/7}$
for cylindrical droplets \cite{HGW:PRE:04}. de Ruijter \textit{et
al.} \cite{RCO:Lan:99} derived a combined model by including the
energy dissipation mechanisms from both the linearized kinetic and
hydrodynamic models. This model does not consider the viscous dissipation
in the precursor foot. For a cylindrical droplet, the combined spreading
model is \cite{HGW:PRE:04}

\begin{equation}
\frac{dr_{b}(t)}{dt}=\frac{\gamma}{\frac{\zeta_{0}}{2}+\frac{2\eta\left(r_{b}(t)-a\right)\sin^{2}\theta}{r_{b}(t)\left(\theta-\sin\theta\cos\theta\right)}}\left(\frac{\theta}{\sin\theta}-\frac{\theta_{0}}{\sin\theta_{0}}\right),\label{eq:combined}\end{equation}

\begin{equation}
\frac{d\theta}{dt}=\left(\frac{\theta-\sin\theta\cos\theta}{A}\right)^{1/2}\left(\cos\theta-\frac{\sin^{3}\theta}{\theta-\sin\theta\cos\theta}\right)^{-1}\frac{dr_{b}(t)}{dt}.\label{eq:combined2}\end{equation}
 where the cross-sectional area $A\sim R_{0}^{2}$, $\gamma$ is the
liquid/vapor surface tension, $\zeta_{0}$ is the friction coefficient,
$\eta$ is the bulk fluid viscosity, $a$ is the hydrodynamic cutoff,
and $\theta_{0}$ is the equilibrium contact angle. The linearized
kinetic and hydrodynamic models are obtained from Eq. \ref{eq:combined}
by setting $\eta$ and $\zeta_{0}$ to zero, respectively. Although
this produces the linearized version of Blake's kinetic model, experiments
\cite{RCB:Lan:97,RCO:Lan:99,SBG:Lan:99} and simulation \cite{BCC:Lan:97,RBC:Lan:99,HGW:PRE:03,HGW:PRE:04}
have shown that it works quite well. These models are written for
droplets with a finite $\theta_{0}$, but they can also be applied
to wetting droplets by setting $\frac{\theta_{0}}{\sin\theta_{0}}=1$
in Eq. \ref{eq:combined}.

Droplets often spread by extending a precursor foot ahead of the main
droplet. This has been observed in experiments \cite{G:RMP:85,HFC:Nat:89,XSB:PRL:04}
as well as simulation \cite{NAK:PRL:92,BKV:PRL:96,HGW:PRE:03,HGW:PRE:04}.
This foot grows diffusively, though to the best of our knowledge there
are no predictions for the dependence on the droplet size. Droplet
spreading experiments have observed {}``terraced spreading'' where
multiple layers spread on top of the precursor foot. In simulations
of droplet spreading, the precursor foot is present, but terraced
spreading has not been observed. This is probably due to the relatively
short duration of the simulation runs.

Although droplet size effects are often ignored when studying spreading
dynamics, we show that the spreading rate of both the precursor foot
and the bulk change with droplet size. Here, we present MD simulations
for two very different systems, a coarse-grained model of polymer
nanodroplets in the wetting regime and an explicit atom model of Pb
on Cu(111), which is non-wetting, to study the dependence of the spreading
rate of both the droplet and the precursor foot on the initial radius
$R_{0}$ of the droplet. Our results demonstrate that the observed
behavior is a general phenomenon and does not depend on the system
specifics. In both cases, the vapor pressure is low so that spreading
does not occur via vaporization and condensation.

For polymer chains, the polymer is represented by spherical beads
of mass $m$ attached by springs, which interact with a truncated
Lennard-Jones (LJ) potential,

\begin{equation}
U_{LJ}(r)=\left\{ \begin{array}{rl}
4\varepsilon\left[\left(\frac{\sigma}{r}\right)^{12}-\left(\frac{\sigma}{r}\right)^{6}\right] & r\leq r_{c}\\
0 & r>r_{c}\end{array}\right.\label{eq:ljcut}\end{equation}
 where $\varepsilon$ and $\sigma$ are the LJ units of energy and
length and the cutoff $r_{c}=2.5\:\sigma$. The monomer-monomer interaction
$\varepsilon$ is used as the reference and all monomers have the
same diameter $\sigma$. For bonded monomers, we apply an additional
potential where each bond is described by the FENE potential \cite{KG:JCP:90}
with $k=30\:\varepsilon/\sigma^{2}$ and $R_{0}=1.5\:\sigma$. The
substrate is modeled as a flat surface since it was found previously
\cite{HGW:PRE:03} that with the proper choice of thermostat, the
simulations using a flat surface exhibit the same behavior as a realistic
atomic substrate with greater computational efficiency. The interactions
between the surface and the monomers in the droplet at a distance
$z$ from the surface are modeled using an integrated LJ potential
with the cutoff set to $z_{c}=2.2\sigma$ \cite{HGW:PRE:03}. Extending
the range of this surface interaction to infinity increases the spreading
rate of the precursor foot and slightly increases the spreading rate
of the bulk. Aside from shifting the wetting transition to a lower
energy, the qualitative spreading behavior is identical to the $z_{c}=2.2\sigma$
case. Specifically, the absence of terraced wetting is probably not
due to the finite interaction range of the surface used in simulations
but to the length of the simulations. Here we present results for
$\varepsilon_{w}=2.0\,\varepsilon$ which for $N=10$ is above the
wetting transition $\varepsilon_{w}^{c}=1.75\,\varepsilon$.

We apply the Langevin thermostat to provide a realistic representation
of the transfer of energy in the polymer droplet. The Langevin thermostat
simulates a heat bath by adding Gaussian white noise and friction
terms to the equation of motion,

\begin{equation}
m_{i}\frac{d^{2}\mathbf{r_{i}}}{dt^{2}}=-\Delta U_{i}-m_{i}\gamma_{L}\frac{d\mathbf{r_{i}}}{dt}+\mathbf{W}_{i}(t),\label{eq:lang}\end{equation}
 where $m_{i}$ is the mass of monomer $i$, $\gamma_{L}$ is the
friction parameter for the Langevin thermostat, $-\Delta U_{i}$ is
the force acting on monomer $i$ due to the potentials defined above,
and $\mathbf{W}_{i}(t)$ is a Gaussian white noise term. Coupling
all of the monomers to the Langevin thermostat would have the unphysical
effect of screening the hydrodynamic interactions in the droplet and
not damping the monomers near the surface stronger than those in the
bulk. To overcome this, we use a Langevin coupling term with a damping
rate that decreases exponentially away from the substrate \cite{BP:PRE:01}.
We choose the form $\gamma_{L}(z)=\gamma_{L}^{s}\exp\left(\sigma-z\right)$
where $\gamma_{L}^{s}$ is the surface Langevin coupling and $z$
is the distance from the substrate. In this paper, we present results
for $\gamma_{L}^{s}=3.0$ and $10.0\,\tau^{-1}$. The larger $\gamma_{L}^{s}$
corresponds to an atomistic substrate with large corrugation and hence
large dissipation and slower diffusion near the substrate.

For Pb on Cu(111), interactions are described via embedded atom method
(EAM) interatomic potentials wherein the energy for $N$ atoms is
\cite{daw83}\begin{equation}
E=\sum_{i=1}^{N}\big[F_{i}(\rho_{i})+{\frac{1}{2}}\sum_{j\ne i}\phi_{ij}(r)\big].\label{eq:eamsysE}\end{equation}
 In Eq. \ref{eq:eamsysE}, $\rho_{i}$ is the electron density at
atom $i$, $\rho_{i}=\sum_{j\ne i}{\rho_{j}}^{a}(r),$ where ${\rho_{j}}^{a}(r)$
is the spherically symmetric electron density contributed by atom
$j$, a distance $r$ from $i$. $F_{i}(\rho_{i})$ is the energy
associated with embedding atom $i$ into an electron density $\rho_{i}$
and $\phi_{ij}(r)$ is a pair potential between atoms $i$ and $j$.
The many-body nature of $F_{i}(\rho_{i})$ in Eq. \ref{eq:eamsysE}
makes the EAM superior to pair potentials for describing bonding in
metal systems. The interactions for Cu, Pb, and the cross-term between
them were previously parameterized \cite{foiles86,lim92,hoyt03}.
The Cu(111) substrate was described via an explicit atom description
with dimension in the surface normal direction equal to four times
the potential cutoff $r_{c}=5.5\,$\AA. The substrate was equilibrated
at the proper lattice constant prior to joining with the drop. Atoms
in the $2r_{c}$ planes furthest from the surface are held rigid and
the rest are permitted to relax according to MD equations of motion
throughout all simulations. Because the substrate is represented atomistically
for Pb(l) on Cu(111), we thermostat only atoms in the substrate using
a Nose-Hoover thermostat algorithm.

All of the droplets presented here are modeled as hemicylinders as
described previously \cite{HGW:PRE:04}. The system is periodic in
the y direction with length $L_{y}$ and open in the other two directions.
This allows a larger droplet radius to be studied using the same number
of monomers than in the spherical geometry. Polymeric droplets have
initial droplet radii of $R_{0}\cong50,$ $80,$ and $120\,\sigma$,
cross-sectional areas $A\cong2690,$ $7490,$ and $18\,200\,\sigma^{2}$
and a total size $N\cong200\,000$, $350\,000,$ and $780\,000$ monomers,
respectively, with $L_{y}=60\,\sigma$ for $R_{0}\cong50\,\sigma$
and $L_{y}=40\,\sigma$ for $R_{0}\geq80\,\sigma$. Pb(l) droplets
are studied with $R_{0}\cong20$, $30$, and $40\, nm$, $A\cong630,$
$1400,$ and $2500\, nm^{2}$ and $N\cong55\,000,$ $122\,000,$ and
$220\,000$ atoms in the drop, respectively, with $L_{y}=27\,$\AA.
In each case, keeping $L_{y}<R_{0}$ or $L_{y}\approx R_{0}$ supresses
any Rayleigh instability. The substrates for the three drop sizes
in the Pb(l) on Cu(111) systems contained $N\cong85\,000,$ $128\,000,$
and $170\,000$ Cu atoms.

The equations of motion are integrated using a velocity-Verlet algorithm.
For polymer spreading, we use a time step of $\Delta t=0.01\;\tau$
where $\tau=\sigma\left(\frac{m}{\varepsilon}\right)^{1/2}$. The
simulations are performed at a temperature $T=\varepsilon/k_{B}$
using the \textsc{lammps} code \cite{P:JCP:95}. For Pb(l) on Cu(111),
$\Delta t=1\; fs$, $T=700\; K$, and the code \textsc{paradyn} \cite{plimpton93}
was used.

\begin{figure}
\begin{center}\includegraphics[%
  clip,
  width=8cm,
  keepaspectratio]{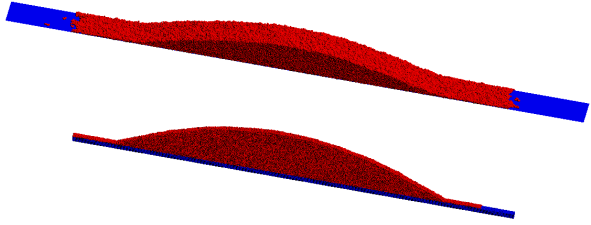}\end{center}

\caption{\label{cap:dropimg} (color online) Cross-section snapshots from
the simulations. A droplet composed of chain length $N=10$ polymers
with initial radius $R_{0}=80\,\sigma$, thickness $L_{y}=40\,\sigma$,
$\varepsilon_{w}=2.0\,\varepsilon$, and $\gamma_{L}^{s}=3.0\,\tau^{-1}$
at $t=71\,600\,\tau$ (top). A Pb(l) droplet with $R_{0}=40\, nm$
and $L_{y}=27\,\textrm{Å}$ on Cu(111) at $t=3.6\, ns$ (bottom).
The snapshots show substrates that are $750\,\sigma$ and $230\, nm$
in length for the polymer and Pb droplets, respectively.}
\end{figure}

For the polymeric system, it was shown previously \cite{HGW:PRE:03}
by measuring the equilibrium contact angle that finite system size
effects become negligible for droplets containing $50\,000$ monomers
or more. Here, we study the size dependence of the spreading behavior
for droplets substantially larger than this finite size limit by using
a minimum droplet size of $200\,000$ monomers in order to simultaneously
study the bulk and precursor foot regions. This is shown in Fig. \ref{cap:dropimg}
which shows the foot extending beyond the bulk region. For the monomeric
liquid system Pb on Cu(111), computational requirements for modeling
the substrate are more significant so we use a minimum drop size $N\sim55\,000$
atoms. In a prior work \cite{webb03}, wetting of Pb(l) drops on Cu(100)
and Cu(111) was simulated using this model. For Pb(l) on Cu(111),
a precursor film rapidly advanced ahead of the main drop in a diffusive
manner (see Fig. \ref{cap:dropimg}); furthermore, the drop reached
an equilibrium but finite contact angle $\theta_{0}=33^{o}$ on top
of the prewetting film, which is in excellent agreement with experiment
\cite{webb03,bailey50,prevot00,moon01}. This system exhibits negligible
exchange of atoms between the liquid and solid, permitting evaluation
of system size effects for non-reactive wetting in the case of a monomeric
liquid with low vapor pressure.

\begin{figure}
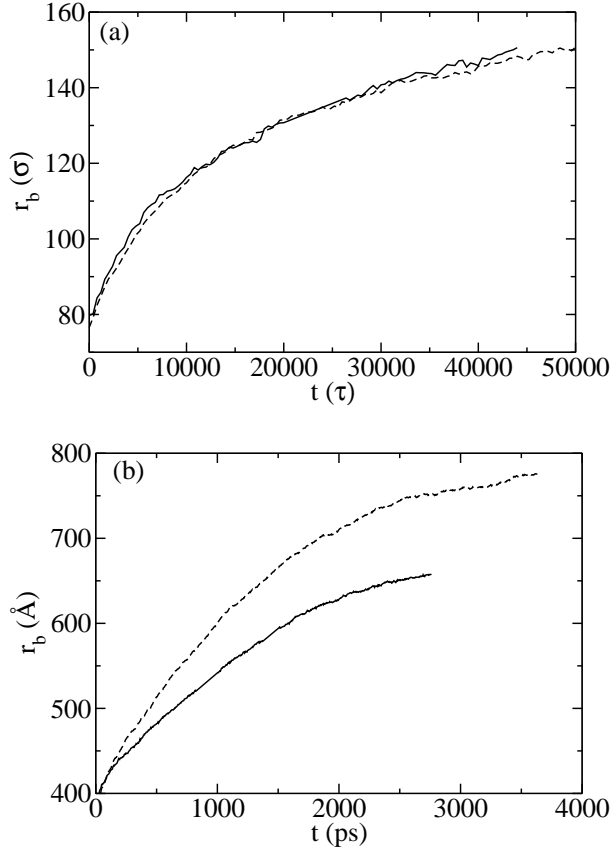

\begin{center}\includegraphics[%
  clip,
  width=8cm,
  keepaspectratio]{fig2a.eps}\end{center}

\begin{center}\includegraphics[%
  clip,
  width=8cm,
  keepaspectratio]{fig2b.eps}\end{center}

\caption{\label{cap:prewet} Comparison of bulk droplet spreading rate on
prewet surface (solid line), and bare surface (dashed line) for (a)
$N=10$ polymers with $\varepsilon_{w}=2.0\,\varepsilon$ and $\gamma_{L}^{s}=10.0\,\tau^{-1}$
and (b) a Pb(l) droplet on Cu(111). }
\end{figure}

For the simulations presented here, we extract the instantaneous contact
radius $r_{b}(t)$ and contact angle $\theta(t)$ every $400\,\tau$
for the polymer system and every $4\, ps$ for the metal system according
to the procedure described previously \cite{HGW:PRE:04,webb03}. To
demonstrate the influence of the viscous dissipation in the precursor
foot on the dynamics of the bulk region, two systems are studied where
the droplets are placed on substrates prewet with a monolayer of the
same material to eliminate the simultaneous spreading of the precursor
foot. The bulk contact radii for both a polymer droplet and a lead
droplet are shown in Fig. \ref{cap:prewet} comparing spreading on
prewet substrates to spreading on bare surfaces. Adding a monolayer
to the substrate does not affect the bulk spreading for the polymeric
system indicating that the viscous dissipation in the precursor foot
is negligible for the droplet \cite{sizenote:}. Although the Pb on
Cu system is drastically different than the polymer system, the results
are very similar. For the lead droplet, the early time behavior shows
that the bulk spreading rate is enhanced as the foot is formed and
begins to extend. Once the foot has extended away from the droplet,
the bulk contact radius curves become parallel. Plotting the velocities
of the precursor foot on both the bare and prewet surfaces shows that
they are identical after one nanosecond.

\begin{figure}
\begin{center}\includegraphics[%
  clip,
  width=8cm,
  keepaspectratio]{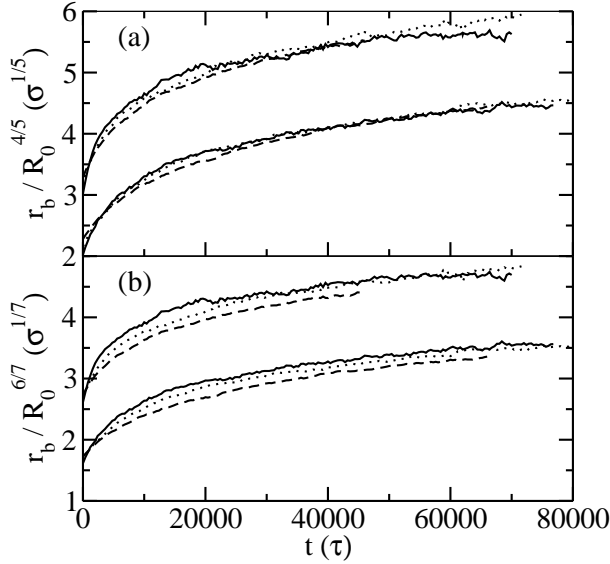}\end{center}

\caption{\label{cap:radbulkscale210} Scaling of the bulk contact radius for
three droplet sizes based on the initial radii $R_{0}=50\,\sigma$
(solid line), $R_{0}=80\,\sigma$ (dotted line), and $R_{0}=120\,\sigma$
(dashed line) using the predictions of (a) the kinetic model and (b)
the hydrodynamic model. Each droplet is composed of chain length $N=10$
polymers with $\varepsilon_{w}=2.0\,\varepsilon$, $\gamma_{L}^{s}=3.0$
(upper curves) and $10.0\,\tau^{-1}$ (lower curves). The $\gamma_{L}^{s}=3.0\,\tau^{-1}$
curves are shifted upward for clarity.}
\end{figure}

\begin{figure}
\begin{center}\includegraphics[%
  clip,
  width=8cm,
  keepaspectratio]{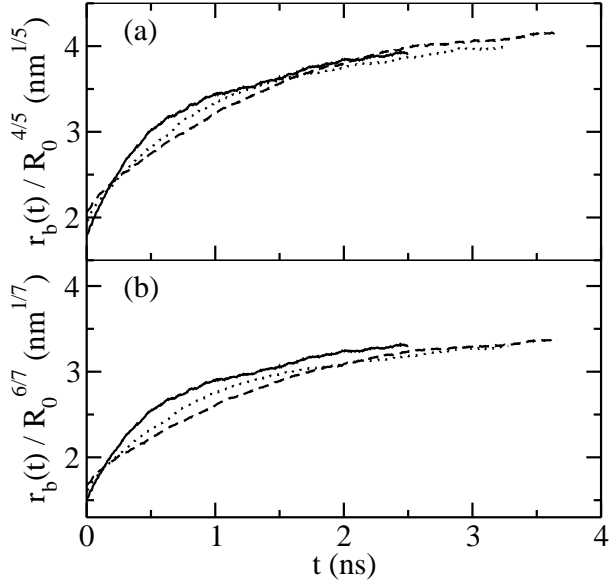}\end{center}

\caption{\label{cap:radscalepb} Scaling of the bulk contact radius of three
Pb(l) droplet sizes based on the initial radii $R_{0}=20\, nm$ (solid
line), $R_{0}=30\, nm$ (dotted line), and $R_{0}=40\, nm$ (dashed
line) using the predictions of (a) the kinetic model and (b) the hydrodynamic
model.}
\end{figure}

The scaling predictions of the linearized kinetic and hydrodynamic
models are applied to the bulk contact radius data for $\varepsilon_{w}=2.0\,\varepsilon$
and $\gamma_{L}^{s}=3.0$ and $10.0\,\tau^{-1}$ in Fig. \ref{cap:radbulkscale210}a
where dividing the bulk contact radius by $R_{0}^{4/5}$ for each
of three droplet sizes causes the data to collapse to a single curve.
The $R_{0}^{6/7}$ scaling of the same data, shown in Fig. \ref{cap:radbulkscale210}b,
does not fit as well because hydrodynamic energy dissipation has only
a weak influence on the spreading rate for the conditions used in
these simulations \cite{HGW:PRE:04}. Similar results are found for
Pb on Cu(111) (see Figs. \ref{cap:radscalepb}a and b) because here
we also have a low viscosity, rapidly spreading droplet.

Previous simulations of polymer droplet spreading \cite{HGW:PRE:04}
have shown that the hydrodynamic model, Eqs. \ref{eq:combined} and
\ref{eq:combined2} with $\zeta_{0}=0$, adequately fits the data
only for the higher viscosity (longer chains) and slower spreading
droplets. Here, a consistent improvement of the fit to the hydrodynamic
model is observed for the three droplet sizes in going from the faster
$\gamma_{L}^{s}=3.0\,\tau^{-1}$ to the slower $\gamma_{L}^{s}=10.0\,\tau^{-1}$
conditions as well as going from smaller to larger droplets. Both
the kinetic and combined models fit the data well for all of the droplets
although the combined model tends to produce less reasonable values
of the fitting parameters \cite{HGW:PRE:03}.

\begin{figure}
\begin{center}\includegraphics[%
  clip,
  width=8cm,
  keepaspectratio]{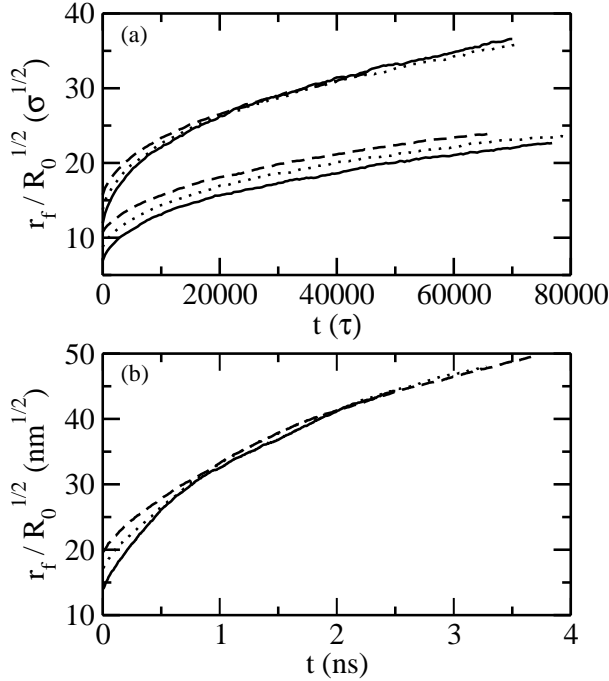}\end{center}

\caption{\label{cap:radfootscale} (a) Polymeric systems precursor foot spreading
rate divided by $R_{0}^{1/2}$ for $R_{0}=50\,\sigma$ (solid line),
$R_{0}=80\,\sigma$ (dotted line), and $R_{0}=120\,\sigma$ (dashed
line) for $\varepsilon_{w}=2.0\,\varepsilon$ and $\gamma_{L}^{s}=10.0\,\tau^{-1}$
(lower curves) and for $\gamma_{L}^{s}=3.0\,\tau^{-1}$ (upper curves).
Each droplet is composed of chain length $N=10$ polymers. The $\gamma_{L}^{s}=3.0\,\tau^{-1}$
curves have been shifted upward for clarity. (b) Precursor foot spreading
rate for Pb(l) on Cu(111) divided by $R_{0}^{1/2}$ for $R_{0}=20,$
$30,$ and $40\, nm$ at $T=700\, K$.}
\end{figure}

The spreading of the precursor foot has been measured experimentally
by ellipsometry and more recently by atomic force microscopy \cite{XSB:PRL:04}.
These studies report effective diffusion coefficients for the precursor
foot without considering the dependence on the droplet size. Joanny
and de Gennes predicted the height profile of the precursor foot to
be proportional to $1/r$ \cite{J:the:85,G:RMP:85}, but models relating
the precursor foot dynamics to the droplet dimensions, such as those
presented above for the bulk dynamics, are not available. Like the
results of the bulk contact radius, the size dependence is evident
in the precursor foot contact radius as well. This is shown in Fig.
\ref{cap:radfootscale} where the precursor foot contact radius is
divided by the initial contact radius raised to the power $n$ where
$n=1/2$. For the $\gamma_{L}^{s}=10.0\,\tau^{-1}$ system, the curves
in Fig. \ref{cap:radfootscale}a show the same asymptotic behavior,
but differences in the initial contact radius cause the curves to
be offset by a constant value at later times. For the $\gamma_{L}^{s}=3.0\,\tau^{-1}$
system, Fig. \ref{cap:radfootscale}a shows the offset is less severe
and the curves for the three different droplet sizes overlap. For
Pb(l) on Cu(111), Fig. \ref{cap:radfootscale}b shows results in this
system are quite similar to what is seen for the $\gamma_{L}^{s}=3.0\,\tau^{-1}$
case in the polymeric systems. This implies that Pb(l) on Cu(111)
corresponds to a lower surface corrugation which, given the significant
lattice mismatch between Pb and Cu and the high density packing of
the (111) surface, is a reasonable result.

The precursor foot contact radius follows $r_{f}^{2}(t)=2D_{eff}t$
where $D_{eff}$ is the effective diffusion coefficient. Within the
uncertainty of our data, these results are consistent with $D_{eff}\sim R_{0}^{x}$
where $x=0.5\pm0.05$ even though the polymeric systems are completely
wetting and the metal system is nonwetting. For the $\gamma_{L}^{s}=10.0\,\tau^{-1}$
system, $D_{eff}\sim R_{0}^{0.65}$ collapses the data onto a master
curve better than $R_{0}^{1/2}$. This may be due to the fact that
the high coupling constant reduces the foot spreading rate to that
of the bulk droplet, so the $D_{eff}\sim R_{0}^{1/2}$ scaling is
not valid when the spreading of the bulk droplet interferes with the
precursor foot diffusion. At present, there is no theoretical model
we are aware of which predicts the dependence of $D_{eff}$ on $R_{0}$,
but it may be because the higher bulk spreading rate of the larger
droplets pushes the precursor foot outward adding to the driving force
of the surface interaction.

In summary, we study the droplet size dependence of nanodroplets spreading
in cylindrical geometries using molecular dynamics simulation. Our
results follow the kinetic model of droplet spreading, which predicts
a $R_{0}^{4/5}$ size scaling of the bulk droplet contact radius.
The bulk spreading rate does not change if the droplet is instead
spread on a prewet surface consisting of a monolayer of the droplet
material. This indicates that, at least in the present simulations,
the viscous dissipation from the precursor foot is not important for
studying the kinetics of the droplet. Theories describing the dynamics
of the precursor foot do not predict a droplet size dependence. The
spreading rate of the precursor foot is known to be diffusive, but
here we show that the effective diffusion coefficient has a droplet
size dependence, $D_{eff}\sim R_{0}^{1/2}$, which can be utilized
in the design of surface wetting applications. For Pb(l) on Cu(111),
use of a realistic interatomic potential results in non-reactive wetting
(in agreement with experiment). Furthermore, vapor pressure is low
despite this being a monomeric liquid. As such, spreading mechanisms
for the metal systems are similar to the polymeric systems studied
and the same size dependence of the spreading rate is found despite
the fact that these are two very different systems.

Sandia is a multiprogram laboratory operated by Sandia Corporation,
a Lockheed Martin Company, for the United States Department of Energy's
National Nuclear Security Administration under Contract No. DE-AC04-94AL85000.

\bibliographystyle{apsrev}
\bibliography{sizedrop}

\begin{thebibliography}{33}
\bibitem[{siz()}]{sizenote:}
\bibinfo{note}{In MD simulations of very small droplets, one can have a
  depletion effect where the drop spreading is effected by the transport of
  material to the precursor foot, but this is not one of the three dissipation
  mechanisms mentioned by de Gennes and is not relevant here.}
\end{thebibliography}

\end{document}